\documentclass[prl,showpacs,a4paper,superscriptaddress,twocolumn]{revtex4}
\usepackage{graphicx,multirow,color,xspace,hyperref}
\usepackage{amsfonts,amsmath}
\usepackage{mathrsfs}
\usepackage[innercaption]{sidecap}
\usepackage{epstopdf}

\begin{document}
%--------------------------------------------------
\setcounter{topnumber}{1}
%--------------------------------------------------
\title{Experimental demonstration of the supersonic-subsonic bifurcation in the circular jump: A hydrodynamic white hole}
%--------------------------------------------------
\author{G. Jannes}
\affiliation{Universit\'{e} de Nice Sophia Antipolis, Laboratoire J.-A. Dieudonn\'{e}, UMR
CNRS-UNS 6621, Parc Valrose, 06108 Nice Cedex 02, France}
%--------------------------------------------------------------
\author{R. Piquet}
\affiliation{Universit\'{e} de Nice Sophia Antipolis, Laboratoire J.-A. Dieudonn\'{e}, UMR CNRS-UNS 6621, Parc Valrose, 06108 Nice Cedex 02, France}
\affiliation{Universit\'{e} Pierre et Marie Curie Paris VI, 4 Place Jussieu, 75005 Paris, France}
%--------------------------------------------------------------
\author{P. Ma\"issa}
\affiliation{Universit\'{e} de Nice Sophia Antipolis, Laboratoire J.-A. Dieudonn\'{e}, UMR
CNRS-UNS 6621, Parc Valrose, 06108 Nice Cedex 02, France}
%--------------------------------------------------------------
\author{C. Mathis}
\affiliation{Universit\'{e} de Nice Sophia Antipolis, Laboratoire J.-A. Dieudonn\'{e}, UMR
CNRS-UNS 6621, Parc Valrose, 06108 Nice Cedex 02, France}
%--------------------------------------------------------------
\author{G. Rousseaux\footnote{Germain.Rousseaux@unice.fr}}
\affiliation{Universit\'{e} de Nice Sophia Antipolis, Laboratoire J.-A. Dieudonn\'{e}, UMR
CNRS-UNS 6621, Parc Valrose, 06108 Nice Cedex 02, France}
%--------------------------------------------------------------
\date{\today}
%--------------------------------------------------

%--------------------------------------------------------------
\begin{abstract}
We provide an experimental demonstration that the circular hydraulic jump represents a hydrodynamic white hole or gravitational fountain (the time-reverse of a black hole) by measuring the angle of the Mach cone created by an object in the ``supersonic'' inner flow region. We emphasise the general character of this gravitational analogy by showing theoretically that the white hole horizon constitutes a stationary and spatial saddle-node bifurcation within dynamical-systems theory. 
We also demonstrate that the inner region has a ``superluminal'' dispersion relation, i.e., that the group velocity of the surface waves increases with frequency, and discuss some possible consequences with respect to the robustness of Hawking radiation. Finally, we point out that our experiment shows a concrete example of a possible ``transplanckian distortion'' of black/white holes.
\end{abstract}
%-----------------------------------------------------------------------
\pacs{47.15.gm, 47.55.N-, 47.40.Ki, 04.70.-s, 04.80.Cc}
\maketitle
A vertical fluid jet impacting on a horizontal plate forms, within a wide range of parameters, a thin layer which expands radially and is surrounded by a sudden circular hydraulic jump. The first modern description of this phenomenon dates back to Lord Rayleigh~\cite{Rayleigh:1914}, who developed a momentum-balance theory to describe it, but did not take viscosity into account. The standard theory for viscous fluids is due to Watson~\cite{Watson:1964}, and has been further improved through the inclusion of surface tension by Bush and Aristoff~\cite{Bush:2003}.
The circular jump is an intricate phenomenon of fluid dynamics: while it suffices to open a kitchen tap to observe it, the theory describing it becomes tremendously complicated for all except its most simple applications. For example, the appearance of more exotic forms such as polygones~\cite{Ellegaard:1998}, through variations of the surface tension~\cite{Bush:2006} or when the liquid flows over micro-textured surfaces~\cite{Stone:2009}, has been studied experimentally but a solid understanding at the theoretical level is still in its infancy. Even for the simple circular jump in a viscous fluid with non-negligible surface tension, 
predictions for the jump radius based on the standard Watson-Bush theory or from a recent and more general description based on lubrication theory~\cite{Rojas:2010} lead to rather involved expressions which at small flow rates only approximately agree with experiments.

Here, we are concerned with a surprising application of the circular hydraulic jump with respect to some of the most exotic objects thought to populate our universe: black holes. Indeed, the circular jump is assumed to constitute an effective white hole (the time-reverse of a black hole) for waves propagating at a speed $c$ on the surface of the fluid ($c=\sqrt{gh}$ in the shallow-water gravity wave limit, with $h$ the fluid height and $g$ the gravitational constant). Theoretically, it is hypothesised (following Rayleigh) that the flow decelerates across the jump from a supercritical flow in the inner region---where the radial fluid velocity at the surface $v_r^s$ is such that $v_r^s>c$ so that surface ripples can only propagate downstream---to a subcritical flow outside, where $v_r^s<c$ and hence the ripples can propagate in both directions. Here, supercritical and subcritical typically refer to the value of the Froude (or Mach) number $Fr=v_r^s/c$. The jump would therefore constitute a one-directional membrane or white hole: surface waves outside the jump cannot penetrate in the inner region; they are trapped outside in precisely the same sense as light is trapped inside a black hole. This analogy can formally be written in relativistic language, as demonstrated by Unruh and Sch\"utzhold for long gravity waves effectively propagating in one dimension~\cite{Schutzhold:2002rf} and applied to the circular jump by Volovik~\cite{Volovik:2005ga,Volovik:2006cz}. The essential point is similar to the case of acoustic black holes~\cite{Unruh:1980cg} and other examples of analogue gravity~\cite{Barcelo:2005fc}: the propagation of these surface waves obeys a generalised d'Alembertian equation in which the intervening curved-spacetime metric is identical (modulo a---physically  irrelevant---global prefactor $1/c^2$) to the 2+1-dimensional Painlev\'{e}-Gullstrand-Lema\^{i}tre (PGL) form of the well-known Schwarzschild metric which describes black/white holes in relativity~\cite{Ray:2007}. Indeed, the line element for the circular jump is
\begin{equation}
ds^2=\frac{1}{c^2}\left[[c^2-(v_r^s)^2]dt^2+ 2v_r^s dt\, dr -dr^2-r^2d\phi^2\right],
\end{equation}
where the surface wave propagation speed $c$ plays the role of the speed of light in gravity, while the radial surface flow velocity $v_r^s$ corresponds to the local velocity of a freely falling observer in the case of gravity. The PGL metric
\begin{equation}
g_{\mu\nu}=
\begin{pmatrix}
c^2-(v_r^s)^2&& v_r^s\\v_r^s&&-I
\end{pmatrix}
\end{equation}
with $I$ the unit matrix, transforms into the Schwarzschild form through the coordinate transformation $d\tilde{t}=dt+dr\, v_r^s/[c^2-(v_r^s)^2]$. The condition $g_{tt}=0$ for a horizon in the non-rotating case becomes simply $c=v_r^s$. It is in this precise sense that the circular jump is believed to constitute the hydrodynamical analogue of a white hole or ``gravitational fountain''.

However, in spite of nearly a century since Rayleigh's description, an explicit experimental proof that the transition from a supercritical to a subcritical flow occurs precisely at the jump, and that the jump hence constitutes a white hole horizon, has so far not been provided. Two strategies could be pursued to provide such a proof. First, one could measure $v_r^s$ and $c$ separately and compare their values. Some measurements of the surface velocity exist, see e.g.~\cite{Bohr:1996}. But these are rather sparse for the inner region. This is probably due to the high value of the speed of flow inside the jump and the complicated nature of its full profile (which could also have an important non-radial component). Moreover, the extreme thinness of the fluid film, typically thinner than a Particle Image Velocimetry laser sheet, means that such imagery methods should be handled with care. Even more complicated is the measure of the surface wave propagation velocity $c$. A direct measure could be performed by sending and tracking surface waves. Possible complications include the dispersive nature of $c$, as well as the ``backreaction'' problem well-known in gravity, i.e., the influence of the wave itself on the geometry of the jump. Alternatively, one could measure the height $h$ of the fluid and in principle derive $c$ as $c=\sqrt{gh}$ plus possible dispersive corrections. 
But this induces an additional approximation which one would prefer to avoid. From the point of view of the white hole analogy, a second and better strategy is therefore to measure the ratio $v_r^s/c$ directly. Our demonstration relies on the Mach cone associated with the supercritical flow in the inner region of the jump.

It is well known that the envelope of the subsequent wavefronts emitted by an object moving at a supersonic speed forms an observable cone, the Mach cone~\cite{Landau:fluid-mechanics}. The half-angle  $\theta$ (or Mach angle) of the cone  can be related to the speed of sound $c_\text{so}$ and the propagation velocity $v$ of the object through simple trigonometry: $\sin\theta=c_\text{so}/v=1/M$ with $M$ the Mach number. The same holds true for a point-like object at rest on the surface of a supercritical fluid flow ($v_r^s>c$) with $c$ now the propagation speed of surface waves.
Measurements of the Mach angle therefore allow to trace the ratio $v_r^s/c$ in the supercritical region. $\theta$ should exactly equal $\pi/2$ at the hydrodynamic white hole horizon where $c=v_r^s$, and become complex (the Mach cone disappearing) in the subcritical region.

\begin{figure}
\includegraphics[width=.48\textwidth, height=52ex]{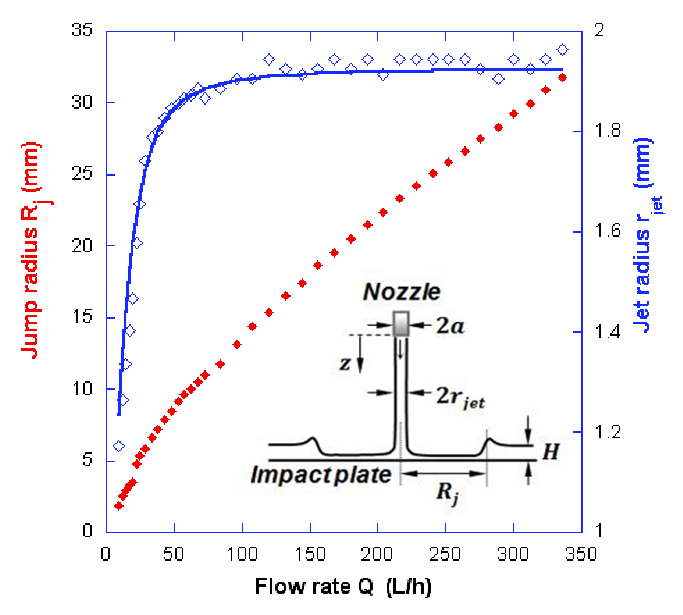}
%\bigskip
\caption{\label{Fig:jump_radius} 
(Color online) Dependence of the jump radius $R_j$ (red filled dots) and the fluid jet radius $r_\text{jet}$ (blue line: theoretical curve, blue hollow diamonds: experimental values measured at a distance $z=13$mm downward from the nozzle) on the flow rate $Q$. Experimental parameters: distance nozzle--impact plate $d$=76mm, nozzle radius $a$=1.925mm, external fluid height far from the jump $H$=0mm. Inset: cross-section diagram of the circular jump.
}
\end{figure}

\emph{Experiments --} Our experiment to demonstrate the presence of a hydrodynamic horizon consists essentially of the following. 
Silicon oil was pumped from within an aquarium through a steel nozzle and impacted on a PVC plate placed inside the aquarium. 
The silicon oil has a high viscosity ($\nu = 20 \text{cS} \approx 20 \nu_\text{water}$), a low surface tension ($\gamma=0.0206\text{N/m}\approx\frac{1}{3}\gamma_\text{water}$), and a density $\rho =950 \text{kg}/\text{m}^3$. The high viscosity allows to maximise the laminarity of the flow. It also guarantees that we create type I circular jumps (with a smooth unidirectional surface flow) over a larger range of flow rates rather than type II jumps (which exhibit surface flow reversal currents near the jump radius) or even turbulent jumps like in water, see e.g.~\cite{Bush:2003}. Such effects might be interesting from a fluid mechanics' point of view, but are detrimental to the gravitational analogy, which assumes a smooth propagation of the surface waves.
A low surface tension moreover guarantees that we avoid polygonal or more exotic jump shapes~\cite{Bush:2006}. The jump that we obtain therefore corresponds to the most straightforward white hole analogy, namely the circularly symmetric (non-rotating) white hole. Fig.~\ref{Fig:jump_radius} shows a typical example of the jump radius $R_j$ and the radius of the fluid jet $r_\text{jet}$ versus flow rate $Q$. The theoretical $r_\text{jet}$-curve is obtained through $r_\text{jet}/a=(1+2gz\pi^2a^4/Q^2)^{-1/4}$, with $a$ the nozzle radius and $z$ the downward distance from the nozzle. 

%=============================================================
%=============================================================

\begin{figure}
\includegraphics[width=.35\textwidth,height=0.25\textwidth]{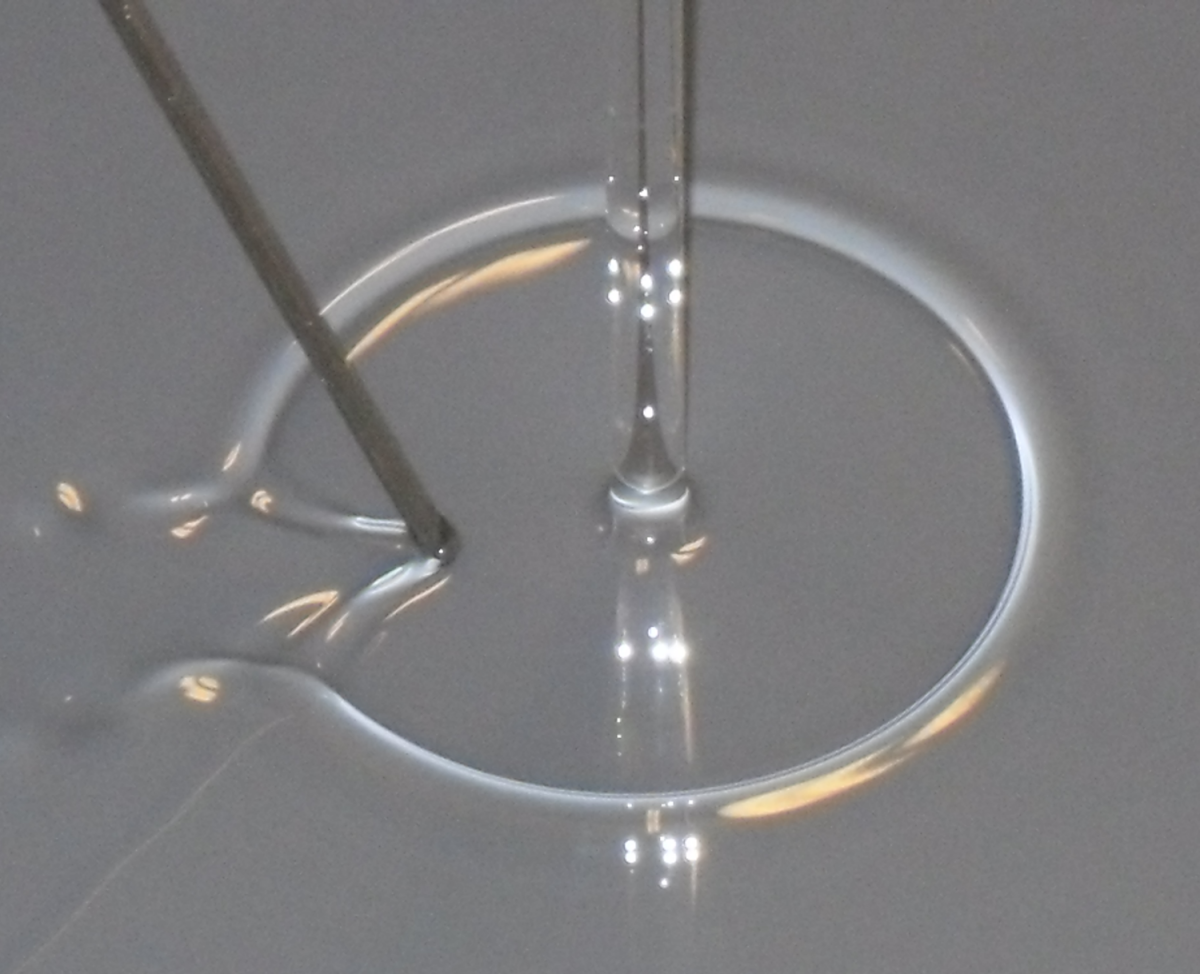}
\caption{\label{Fig:mach-cone1} 
Mach cone in a circular hydraulic jump.}
\end{figure}

\begin{figure}
\includegraphics[width=.30\textwidth]{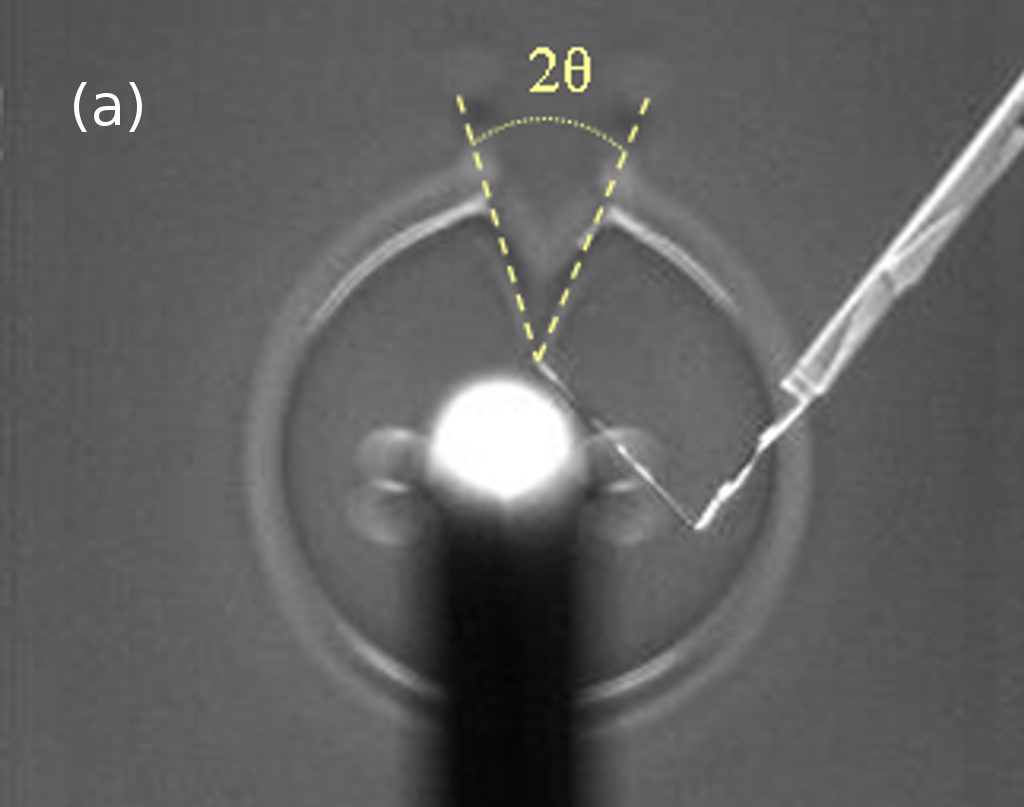}\\

\includegraphics[width=.30\textwidth]{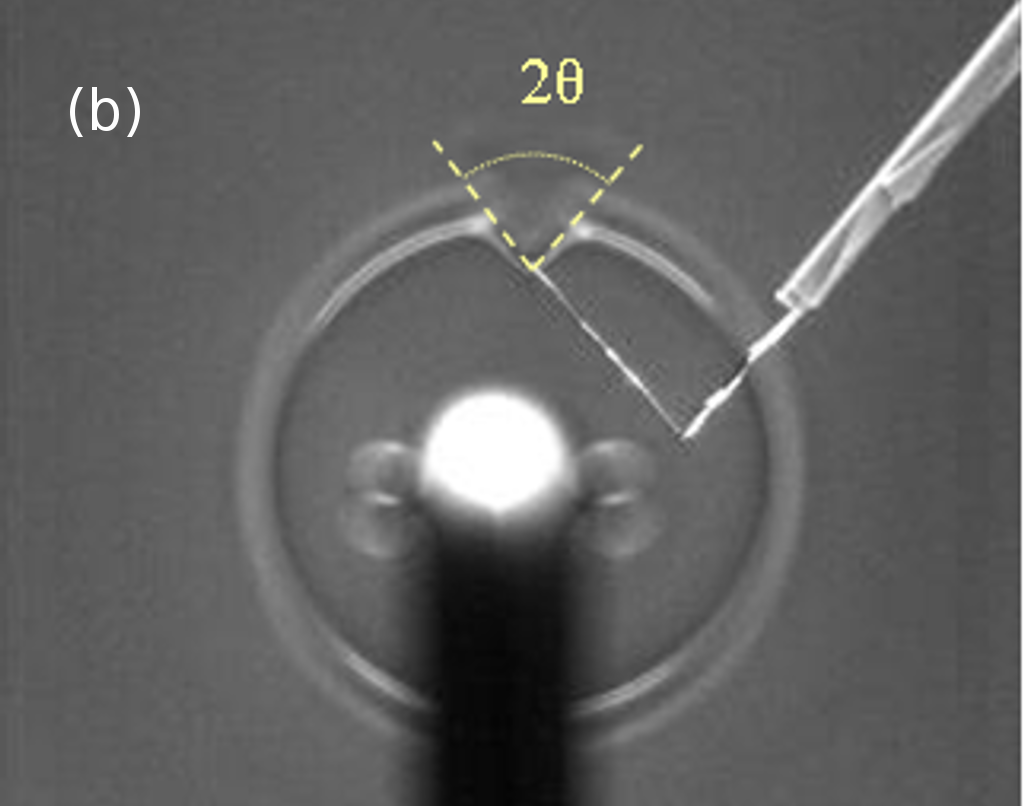}\\

\includegraphics[width=.30\textwidth]{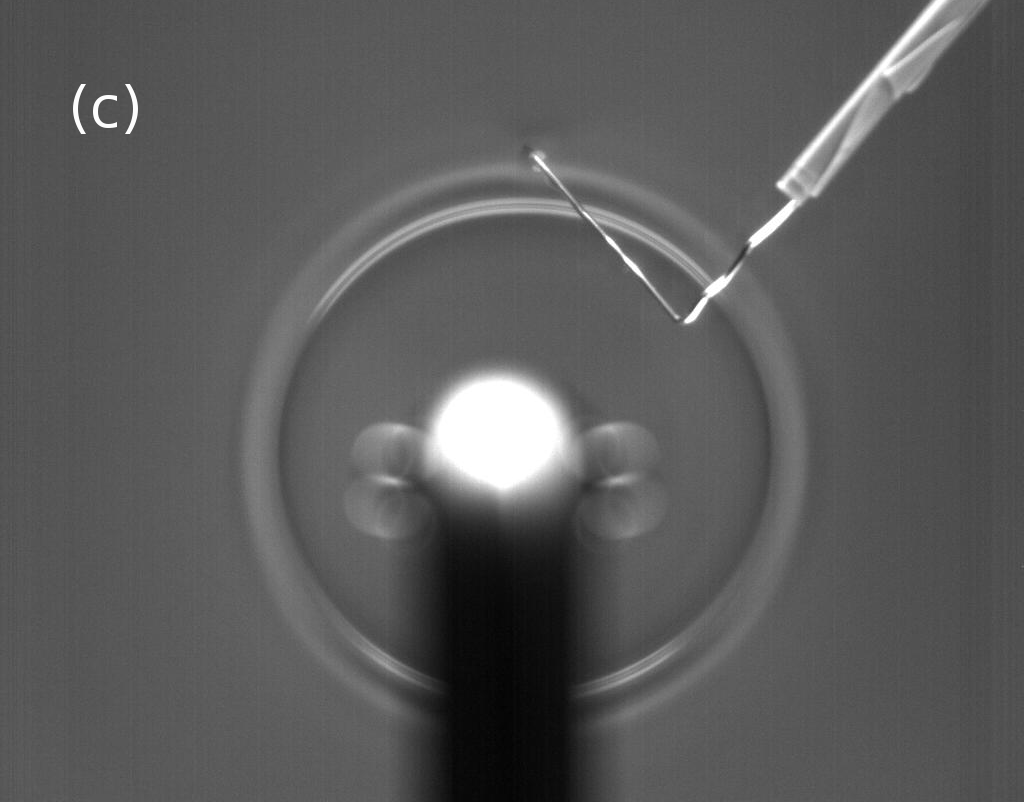}
%\bigskip
\caption{\label{Fig:mach-cone2} 
Measurements of the Mach angle $\theta$ from pictures taken with an overhead camera. A needle is placed inside the flow at varying distances from the centre of the jump. (a) Mach cone near the centre of the jump. (b) Mach cone near the edge of the jump. (c) The Mach cone disappears just outside the jump. [The blurry object in the bottom part of the pictures is the nozzle holder.]}
\end{figure}

\begin{figure}
\includegraphics[width=.48\textwidth,height=52ex]{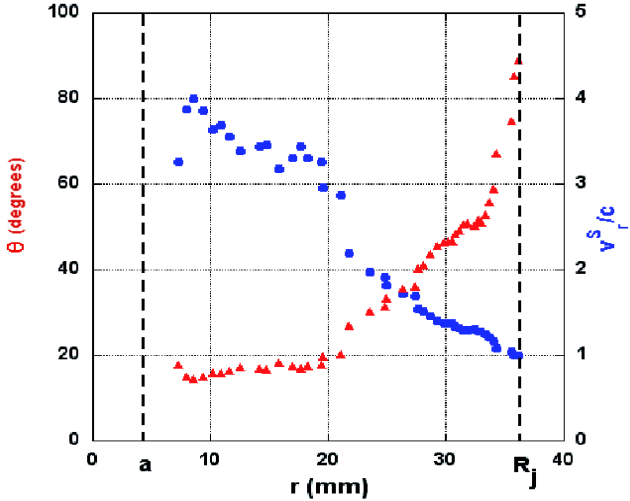}
%\bigskip
\caption{\label{Fig:mach-angle} 
(Color online) Mach angle $\theta$ (red triangles) and ratio $v_ r^s/c$ (blue circles) as a function of the distance $r$ from the centre of the jump. The dashed vertical lines represent the nozzle radius $a$ and the jump radius $R_j$. Experimental parameters:  external fluid height $H=0$mm; nozzle radius $a=4.75$mm; distance nozzle--impact plate $d=62.5$mm; flow rate $Q=240$L/h.} 
\end{figure}

A needle penetrates the flow surface at varying  distance from the jet's impact point (the centre of the circular jump), see Figs.~\ref{Fig:mach-cone1} and~\ref{Fig:mach-cone2}, to create the Mach cones.
We measure the Mach angle as close as possible to the needle. Our main results, the Mach angles $\theta$ as well as the resulting relation $v_r^s/c$, are presented in Fig.~\ref{Fig:mach-angle}. We have checked that these do not depend qualitatively on $H,a, d$ or $Q$, as long as one remains within a stable type I jump regime.
Inside the jet impact zone ($r<a$), we expect $v_r^s\ll c$ followed by a steep increase for $r\gtrsim a$ until a certain value $v_r^{s, \text{max}}$, since the fluid impacts vertically before being converted into a radial flow. The field of vision of our experimental setup starts near this maximum, see Fig.~\ref{Fig:mach-angle}, corresponding to a Mach angle $\theta$ of roughly $\pi/10$. From there, $\theta$ smoothly increases to about $\pi/4$ at approximately 3/4 of the jump radius, and then rapidly opens up to reach exactly $\pi/2$ near the ridge of the jump. This implies $v_r^s/c=1$ and hence
constitutes a clear proof that the jump indeed represents a white hole horizon for surface waves, independently of whether $c$ is strictly equal to $\sqrt{gh}$ or modified by dispersive corrections. Converting the Mach angle into $v_r^s/c$ shows that the latter ratio decreases mainly far from the jump, well inside the inner region. The critical point $v_r^s/c=1$ is actually reached in a very smooth way. This is in curious contrast with the standard theoretical models in fluid mechanics, which describe the circular jump as a shock wave and therefore prescribe that the critical point itself should lie within a sharp (and in models without viscosity: discontinuous) transition from a supercritical to a subcritical regime, see e.g.~\cite{Bohr:1997}. It is not clear whether this smooth transition to the critical point is a genuine property of the jump itself, or a consequence of the perturbation of the flow pattern due to the insertion of the needle, which somehow smoothens out the shock wave (or shifts its position). It should be noted that a perturbation of the flow pattern can indeed be identified with the naked eye when the needle penetrates the flow all the way down to the plate. However, we have precisely carefully avoided this by having the needle penetrate only the surface of the flow. In case this smooth transition is a genuine property of the jump, then the expected sharp transition in the velocity profile might still occur just beyond the critical point. This could suggest that the jump is actually a consequence of the existence of a horizon, rather than vice versa. If the smoothness of the transition is due to a perturbation of the original flow pattern (and hence unavoidable in the kind of experiment that we have carried out), then this would constitute an example of a ``transplanckian'' distortion and reconstruction, of which we discuss a more evident example in the final section of this manuscript.

In any case, it is particularly striking that this white hole horizon can be identified with the naked eye, even without observing the actual blocking of surface waves. Moreover, contrarily to most other examples of analogue gravity, the horizon forms rather spontaneously, without extraordinary engineering.

%=============================================================

%=============================================================

%=============================================================
%=============================================================

\emph{Saddle-node bifurcation -- } The fact that the circular jump represents a white hole horizon illustrates that the concept of horizons is not limited to relativity. This generality goes even further. Note that the Mach angles $\theta_W\epsilon [0,\pi/2]$ and $\theta_B=\pi-\theta_W$ would lead to the same value of $v_r^s/c$, $\theta_W$ corresponding to a white hole (a source) and $\theta_B$ to a black hole (a drain). At the horizon itself, both solutions merge: $\theta_W=\theta_B=\pi/2$. This is an example of a saddle-node bifurcation in dynamical systems theory, as was earlier established in the deep-water gravity-wave regime~\cite{Rousseaux:2009}. Indeed, using an asymptotic development of $\arcsin(x)$ for $x\to 1$, one can write in the near-horizon region (for shallow-water waves and also e.g.\ for acoustic waves in a de Laval nozzle): \mbox{$ \theta=\arcsin(1-\epsilon)\approx \pi/2(1-\sqrt{\epsilon})$}, where $\epsilon=1-c/v_r^s$. Inverting this relation, one obtains 
\begin{equation}
 \epsilon-\left(\frac{\pi/2-\theta}{\pi/2}\right)^2=0.
\end{equation}
This is precisely the canonical expression of the stationary normal form for a spatial saddle-node bifurcation, with $\epsilon$ the control parameter and $\theta$ the order parameter~\cite{Iooss}. It implies that the near-horizon behaviour inside a black/white hole is not only a common feature of typical analogue gravity systems involving sound or surface waves, but belongs to the universality class of saddle-node bifurcations in a dynamical-systems description.

\begin{figure}
\includegraphics[width=.45\textwidth, height=35ex]{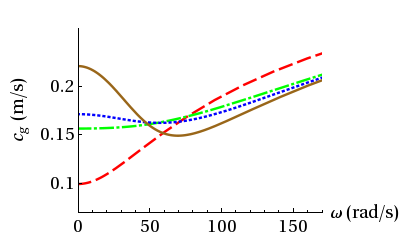}
%\bigskip
\caption{\label{Fig:group-velocities} 
(Color online) Group velocities (at $U=0$) for different values of the fluid height $h$. [From bottom to top at low $\omega$:] $h=1$mm (red dashed line); 2.5mm (green dashed-dotted line); 3mm (blue dotted line); 5mm (full brown line).
}
\end{figure}

\emph{Dispersion relation -- }
The dispersion relation for gravity-capillary surface waves is \mbox{($\omega-Uk)^2=(gk+\frac{\gamma}{\rho}k^3)\tanh(kh)$}~\cite{Landau:fluid-mechanics}. 
In the shallow-water limit $kh\ll 1$, one can tentatively write 
\begin{eqnarray}\label{dispersion-full}
(\omega - Uk)^2 = c^2k^2 + c^2\left(l_c^2-\frac{h^2}{3}\right)k^4 + \mathcal O(k^6).
\end{eqnarray}
with $l_c=\sqrt{\frac{\gamma}{\rho g}}$ the capillary length. At low wave numbers $k$, a relativistic dispersion is indeed recovered. For intermediate $k$'s, the dispersion can either be normal or anomalous---in relativistic language: ``subluminal'' or ``superluminal'' (i.e., the group velocity $c_g\equiv \frac{d\omega}{dk}$ decreases or increases with $k$, respectively), with a critical transition depth $h_\text{trans}= \sqrt{3}l_c$ (see also~\cite{Visser:2007du}), corresponding to an inflection point $d^2c_g/dk^2=0$ at $k=0$. For our silicon oil, $h_\text{trans}\approx 2.6mm$, see Fig.~\ref{Fig:group-velocities}. 
At higher $k$'s,~\eqref{dispersion-full} can no longer be trusted, and the dispersion becomes superluminal, irrespective of the value of $h$, asymptoting to $c_g \propto \sqrt{k}$. Within the limits of the stable type I circular jump regime, the inner region depth $h_\text{in}$ turns out to be always on the order of 1mm in our experiments, and hence $h_\text{in}<h_\text{trans}$. For the type I jump, $h_\text{in}$ is highly insensitive to the external fluid height $H$ which can be imposed artificially, and varies only slightly with the flow rate, in accordance with earlier observations with other fluids~\cite{Bohr:1996,Craik:1981}. Likewise, in water, the fluid depth in the inner region is typically $\lesssim0.5mm$~\cite{Craik:1981}, while the transition height which marks the appearance of a subluminal frequency range is $h_\text{trans}\approx 4.7mm$. We conclude that the inner region of the circular jump naturally exhibits a superluminal dispersion relation. 

%
%
%=============================================================
%=============================================================
%
%

When increasing the external fluid height $H$, care must be taken not to destabilise the type I jump (or drown it altogether). We have verified that $H$ can indeed be increased beyond $h_\text{trans}$ within the limits of the white hole analogy. The outer region can therefore be tuned to be in a super- or in a subluminal regime by varying the external fluid height. This opens several interesting prospects. First, it suggests a new scenario for quantum gravity phenomenology, in which the homogeneity of the dispersion relation is broken and quantum-gravitational effects at the horizon would lead to a separation between two regions with different dispersion. Second, it means that a stimulated Hawking signal (a thermally correlated pair of positive and negative frequency waves emanating from the horizon when this is hit by an incoming wave) should be detectable in the subluminal outside region, much as in~\cite{Weinfurtner:2010nu} (see also~\cite{Rousseaux:2007is,Rousseaux:2010md}). However, a second such pair should also form towards the superluminal interior, see e.g.~\cite{Unruh:2004zk}. 
It is an open question whether these inner and outer pairs should be cross-correlated. A third prospect is the following. When the fluid regime becomes supercritical, friction with a fixed boundary leads to amplification of high-$k$ modes (rather than damping, as in the normal $v_r^s<c$ regime). This effect is associated to the occurence of negative energy modes ($\omega-Uk<0$) and is called the Miles or ergoregion instability. If there exists a transplanckian preferred frame in gravity, then friction with respect to this transplanckian ``ether'' could lead to a similar instability. It has been suggested that such a Miles or ergoregion instability could affect the Hawking process by distorting the ``thermal'' balance between negative and positive frequency modes on opposite sides of the horizon~\cite{Unruh:2004zk}, or that the Miles instability could actually become the dominant mechanism of dissipation of the black hole~\cite{Volovik:2005ga}, reducing the Hawking radiation to a theoretical curiosity. Comparison of the inner and outer correlated pairs in the circular jump could give valuable information about this competition and its influence on the robustness of Hawking radiation. 

\emph{Transplanckian distortion --} A related explicit example of a ``transplanckian distortion'' of the white hole is already apparent in our current experiment. The needle represents a transplanckian object in the analogue gravity system, since it is not subject to the effective hydrodynamic gravity and creates its own preferred reference frame. The presence of the needle strongly distorts the white-hole horizon, see Fig.~\ref{Fig:mach-cone1} above. The original horizon (the jump) is completely destroyed along the arc corresponding to the position of the needle, and reconstructed along the Mach cone, slightly opening up towards the exterior, and reconnecting with the original horizon at the jump radius. This reconstruction process is accompanied by two jets of radiation propagating radially outward from the points of reconnection. Similar distortions could be crucial, for example, in brane-cosmology models in which the brane on which we live is intersected by higher-dimensional objects or other branes.

\section*{Acknowledgements}
The authors thank David Brizuela for a useful comment on the dispersion relation, and Luis Garay, Carlos Barcel\'{o} and Grisha Volovik for general discussions and comments. G.J. thanks INSMI, CNRS and the University of Nice for financial support.  This research was supported by the R\'{e}gion PACA (Projet exploratoire HYDRO) and the Conseil G\'en\'eral 06.


\begin{thebibliography}{24}
\expandafter\ifx\csname natexlab\endcsname\relax\def\natexlab#1{#1}\fi
\expandafter\ifx\csname bibnamefont\endcsname\relax
  \def\bibnamefont#1{#1}\fi
\expandafter\ifx\csname bibfnamefont\endcsname\relax
  \def\bibfnamefont#1{#1}\fi
\expandafter\ifx\csname citenamefont\endcsname\relax
  \def\citenamefont#1{#1}\fi
\expandafter\ifx\csname url\endcsname\relax
  \def\url#1{\texttt{#1}}\fi
\expandafter\ifx\csname urlprefix\endcsname\relax\def\urlprefix{www.arxiv.org}\fi
\providecommand{\bibinfo}[2]{#2}
\providecommand{\eprint}[2][]{\url{#2}}

\bibitem[{\citenamefont{Rayleigh}(1914)}]{Rayleigh:1914}
\bibinfo{author}{\bibfnamefont{L.}~\bibnamefont{Rayleigh}},
  \bibinfo{journal}{Proc. Roy. Soc. A} \textbf{\bibinfo{volume}{90}},
  \bibinfo{pages}{324} (\bibinfo{year}{1914}).

\bibitem[{\citenamefont{Watson}(1964)}]{Watson:1964}
\bibinfo{author}{\bibfnamefont{E.~J.} \bibnamefont{Watson}},
  \bibinfo{journal}{J. Fluid Mech.} \textbf{\bibinfo{volume}{20}},
  \bibinfo{pages}{481} (\bibinfo{year}{1964}).

\bibitem[{\citenamefont{Bush and Aristoff}(2003)}]{Bush:2003}
\bibinfo{author}{\bibfnamefont{J.~W.~M.} \bibnamefont{Bush}} \bibnamefont{and}
  \bibinfo{author}{\bibfnamefont{J.~M.} \bibnamefont{Aristoff}},
  \bibinfo{journal}{J. Fluid. Mech.} \textbf{\bibinfo{volume}{489}},
  \bibinfo{pages}{229} (\bibinfo{year}{2003}).

\bibitem[{\citenamefont{Ellegaard et~al.}(1998)\citenamefont{Ellegaard, Hansen,
  Haaning, Marcussen, Bohr, Hansen, and Watanabe}}]{Ellegaard:1998}
\bibinfo{author}{\bibfnamefont{C.}~\bibnamefont{Ellegaard}},
  \bibinfo{author}{\bibfnamefont{A.~E.} \bibnamefont{Hansen}},
  \bibinfo{author}{\bibfnamefont{A.}~\bibnamefont{Haaning}},
  \bibinfo{author}{\bibfnamefont{A.}~\bibnamefont{Marcussen}},
  \bibinfo{author}{\bibfnamefont{T.}~\bibnamefont{Bohr}},
  \bibinfo{author}{\bibfnamefont{J.}~\bibnamefont{Hansen}}, \bibnamefont{and}
  \bibinfo{author}{\bibfnamefont{S.}~\bibnamefont{Watanabe}},
  \bibinfo{journal}{Nature} \textbf{\bibinfo{volume}{392}},
  \bibinfo{pages}{767} (\bibinfo{year}{1998}).

\bibitem[{\citenamefont{Bush et~al.}(2006)\citenamefont{Bush, Aristoff, and
  Hosoi}}]{Bush:2006}
\bibinfo{author}{\bibfnamefont{J.~W.~M.} \bibnamefont{Bush}},
  \bibinfo{author}{\bibfnamefont{J.~M.} \bibnamefont{Aristoff}},
  \bibnamefont{and} \bibinfo{author}{\bibfnamefont{A.~E.} \bibnamefont{Hosoi}},
  \bibinfo{journal}{J. Fluid. Mech.} \textbf{\bibinfo{volume}{558}},
  \bibinfo{pages}{33} (\bibinfo{year}{2006}).

\bibitem[{\citenamefont{Dressaire et~al.}(2009)\citenamefont{Dressaire,
  Courbin, Crest, and Stone}}]{Stone:2009}
\bibinfo{author}{\bibfnamefont{E.}~\bibnamefont{Dressaire}},
  \bibinfo{author}{\bibfnamefont{L.}~\bibnamefont{Courbin}},
  \bibinfo{author}{\bibfnamefont{J.}~\bibnamefont{Crest}}, \bibnamefont{and}
  \bibinfo{author}{\bibfnamefont{H.~A.} \bibnamefont{Stone}},
  \bibinfo{journal}{Phys. Rev. Lett.} \textbf{\bibinfo{volume}{102}},
  \bibinfo{pages}{194503} (\bibinfo{year}{2009}).

\bibitem[{\citenamefont{Rojas et~al.}(2010)\citenamefont{Rojas, Argentina,
  Cerda, and Tirapegui}}]{Rojas:2010}
\bibinfo{author}{\bibfnamefont{N.~O.} \bibnamefont{Rojas}},
  \bibinfo{author}{\bibfnamefont{M.}~\bibnamefont{Argentina}},
  \bibinfo{author}{\bibfnamefont{E.}~\bibnamefont{Cerda}}, \bibnamefont{and}
  \bibinfo{author}{\bibfnamefont{E.}~\bibnamefont{Tirapegui}},
  \bibinfo{journal}{Phys. Rev. Lett.} \textbf{\bibinfo{volume}{104}},
  \bibinfo{pages}{187801} (\bibinfo{year}{2010}).

\bibitem[{\citenamefont{Schutzhold and Unruh}(2002)}]{Schutzhold:2002rf}
\bibinfo{author}{\bibfnamefont{R.}~\bibnamefont{Schutzhold}} \bibnamefont{and}
  \bibinfo{author}{\bibfnamefont{W.~G.} \bibnamefont{Unruh}},
  \bibinfo{journal}{Phys. Rev.} \textbf{\bibinfo{volume}{D66}},
  \bibinfo{pages}{044019} (\bibinfo{year}{2002}), \eprint{arXiv:gr-qc/0205099}.

\bibitem[{\citenamefont{Volovik}(2005)}]{Volovik:2005ga}
\bibinfo{author}{\bibfnamefont{G.~E.} \bibnamefont{Volovik}},
  \bibinfo{journal}{JETP Lett.} \textbf{\bibinfo{volume}{82}},
  \bibinfo{pages}{624} (\bibinfo{year}{2005}), \eprint{arXiv:physics/0508215}.

\bibitem[{\citenamefont{Volovik}(2006)}]{Volovik:2006cz}
\bibinfo{author}{\bibfnamefont{G.~E.} \bibnamefont{Volovik}},
  \bibinfo{journal}{J. Low Temp. Phys.} \textbf{\bibinfo{volume}{145}},
  \bibinfo{pages}{337} (\bibinfo{year}{2006}), \eprint{arXiv:gr-qc/0603093}.

\bibitem[{\citenamefont{Unruh}(1981)}]{Unruh:1980cg}
\bibinfo{author}{\bibfnamefont{W.~G.} \bibnamefont{Unruh}},
  \bibinfo{journal}{Phys. Rev. Lett.} \textbf{\bibinfo{volume}{46}},
  \bibinfo{pages}{1351} (\bibinfo{year}{1981}).

\bibitem[{\citenamefont{Barcel\'{o} et~al.}(2005)\citenamefont{Barcel\'{o},
  Liberati, and Visser}}]{Barcelo:2005fc}
\bibinfo{author}{\bibfnamefont{C.}~\bibnamefont{Barcel\'{o}}},
  \bibinfo{author}{\bibfnamefont{S.}~\bibnamefont{Liberati}}, \bibnamefont{and}
  \bibinfo{author}{\bibfnamefont{M.}~\bibnamefont{Visser}},
  \bibinfo{journal}{Living Rev. Rel.} \textbf{\bibinfo{volume}{8}},
  \bibinfo{pages}{12} (\bibinfo{year}{2005}), \eprint{arXiv:gr-qc/0505065}.

\bibitem[{\citenamefont{Ray and Bhattacharjee}(2007)}]{Ray:2007}
\bibinfo{author}{\bibfnamefont{A.~K.} \bibnamefont{Ray}} \bibnamefont{and}
  \bibinfo{author}{\bibfnamefont{J.~K.} \bibnamefont{Bhattacharjee}},
  \bibinfo{journal}{Phys. Lett. A} \textbf{\bibinfo{volume}{371}},
  \bibinfo{pages}{241} (\bibinfo{year}{2007}).

\bibitem[{\citenamefont{Bohr et~al.}(1996)\citenamefont{Bohr, Ellegaard,
  Hansen, and Haaning}}]{Bohr:1996}
\bibinfo{author}{\bibfnamefont{T.}~\bibnamefont{Bohr}},
  \bibinfo{author}{\bibfnamefont{C.}~\bibnamefont{Ellegaard}},
  \bibinfo{author}{\bibfnamefont{A.~E.} \bibnamefont{Hansen}},
  \bibnamefont{and} \bibinfo{author}{\bibfnamefont{A.}~\bibnamefont{Haaning}},
  \bibinfo{journal}{Physica B} \textbf{\bibinfo{volume}{228}},
  \bibinfo{pages}{1} (\bibinfo{year}{1996}).

\bibitem[{\citenamefont{Landau and Lifshitz}()}]{Landau:fluid-mechanics}
\bibinfo{author}{\bibfnamefont{L.~D.} \bibnamefont{Landau}} \bibnamefont{and}
  \bibinfo{author}{\bibfnamefont{E.~M.} \bibnamefont{Lifshitz}},
  \emph{\bibinfo{title}{Fluid Mechanics}}, \bibinfo{note}{2nd ed.,
  Pergamon, UK (1987)}.

\bibitem[{\citenamefont{Bohr et~al.}(1997)\citenamefont{Bohr, Putkaradze, and
  Watanabe}}]{Bohr:1997}
\bibinfo{author}{\bibfnamefont{T.}~\bibnamefont{Bohr}},
  \bibinfo{author}{\bibfnamefont{V.}~\bibnamefont{Putkaradze}},
  \bibnamefont{and} \bibinfo{author}{\bibfnamefont{S.}~\bibnamefont{Watanabe}},
  \bibinfo{journal}{Phys. Rev. Lett.} \textbf{\bibinfo{volume}{79}},
  \bibinfo{pages}{1038} (\bibinfo{year}{1997}).

\bibitem[{\citenamefont{Nardin et~al.}(2009)\citenamefont{Nardin, Rousseaux,
  and Coullet}}]{Rousseaux:2009}
\bibinfo{author}{\bibfnamefont{J.-C.} \bibnamefont{Nardin}},
  \bibinfo{author}{\bibfnamefont{G.}~\bibnamefont{Rousseaux}},
  \bibnamefont{and} \bibinfo{author}{\bibfnamefont{P.}~\bibnamefont{Coullet}},
  \bibinfo{journal}{Phys. Rev. Lett.} \textbf{\bibinfo{volume}{102}},
  \bibinfo{pages}{124504} (\bibinfo{year}{2009}).

\bibitem[{\citenamefont{Haragus and Iooss}()}]{Iooss}
\bibinfo{author}{\bibfnamefont{M.}~\bibnamefont{Haragus}} \bibnamefont{and}
  \bibinfo{author}{\bibfnamefont{G.}~\bibnamefont{Iooss}},
  \emph{\bibinfo{title}{Local Bifurcations, Center Manifolds, and Normal Forms
  in Infinite-Dimensional Dynamical Systems}}, \bibinfo{note}{eDP Sci.
  and Springer (2010)}.

\bibitem[{\citenamefont{Visser and Weinfurtner}(2007)}]{Visser:2007du}
\bibinfo{author}{\bibfnamefont{M.}~\bibnamefont{Visser}} \bibnamefont{and}
  \bibinfo{author}{\bibfnamefont{S.}~\bibnamefont{Weinfurtner}},
  \bibinfo{journal}{PoS} \textbf{\bibinfo{volume}{QG-PH}}, \bibinfo{pages}{042}
  (\bibinfo{year}{2007}), \eprint{arXiv:0712.0427}.

\bibitem[{\citenamefont{Craik et~al.}(1981)\citenamefont{Craik, Latham, Fawkes,
  and Gribbon}}]{Craik:1981}
\bibinfo{author}{\bibfnamefont{A.~D.~D.} \bibnamefont{Craik}},
  \bibinfo{author}{\bibfnamefont{R.~C.} \bibnamefont{Latham}},
  \bibinfo{author}{\bibfnamefont{M.~J.} \bibnamefont{Fawkes}},
  \bibnamefont{and} \bibinfo{author}{\bibfnamefont{P.~W.~F.}
  \bibnamefont{Gribbon}}, \bibinfo{journal}{J. Fluid Mech.}
  \textbf{\bibinfo{volume}{112}}, \bibinfo{pages}{347} (\bibinfo{year}{1981}).

\bibitem[{\citenamefont{Weinfurtner et~al.}(2010)\citenamefont{Weinfurtner,
  Tedford, Penrice, Unruh, and Lawrence}}]{Weinfurtner:2010nu}
\bibinfo{author}{\bibfnamefont{S.}~\bibnamefont{Weinfurtner}},
  \bibinfo{author}{\bibfnamefont{E.~W.} \bibnamefont{Tedford}},
  \bibinfo{author}{\bibfnamefont{M.~C.~J.} \bibnamefont{Penrice}},
  \bibinfo{author}{\bibfnamefont{W.~G.} \bibnamefont{Unruh}}, \bibnamefont{and}
  \bibinfo{author}{\bibfnamefont{G.~A.} \bibnamefont{Lawrence}},
  \bibinfo{journal}{Phys. Rev. Lett.}
  \textbf{\bibinfo{volume}{106}}, \bibinfo{pages}{021302} (\bibinfo{year}{2011}), \eprint{arXiv:1008.1911}.

\bibitem[{\citenamefont{Rousseaux et~al.}(2008)\citenamefont{Rousseaux, Mathis,
  Maissa, Philbin, and Leonhardt}}]{Rousseaux:2007is}
\bibinfo{author}{\bibfnamefont{G.}~\bibnamefont{Rousseaux}},
  \bibinfo{author}{\bibfnamefont{C.}~\bibnamefont{Mathis}},
  \bibinfo{author}{\bibfnamefont{P.}~\bibnamefont{Maissa}},
  \bibinfo{author}{\bibfnamefont{T.~G.} \bibnamefont{Philbin}},
  \bibnamefont{and}
  \bibinfo{author}{\bibfnamefont{U.}~\bibnamefont{Leonhardt}},
  \bibinfo{journal}{New J. Phys.} \textbf{\bibinfo{volume}{10}},
  \bibinfo{pages}{053015} (\bibinfo{year}{2008}), \eprint{arXiv:0711.4767}.

\bibitem[{\citenamefont{Rousseaux et~al.}(2010)\citenamefont{Rousseaux, Maissa,
  Mathis, Coullet, Philbin, and Leonhardt}}]{Rousseaux:2010md}
\bibinfo{author}{\bibfnamefont{G.}~\bibnamefont{Rousseaux}},
  \bibinfo{author}{\bibfnamefont{P.}~\bibnamefont{Maissa}},
  \bibinfo{author}{\bibfnamefont{C.}~\bibnamefont{Mathis}},
  \bibinfo{author}{\bibfnamefont{P.}~\bibnamefont{Coullet}},
  \bibinfo{author}{\bibfnamefont{T.~G.} \bibnamefont{Philbin}},
  \bibnamefont{and}
  \bibinfo{author}{\bibfnamefont{U.}~\bibnamefont{Leonhardt}},
  \bibinfo{journal}{New J. Phys.} \textbf{\bibinfo{volume}{12}},
  \bibinfo{pages}{095018} (\bibinfo{year}{2010}), \eprint{arXiv:1004.5546}.

\bibitem[{\citenamefont{Unruh and Schutzhold}(2005)}]{Unruh:2004zk}
\bibinfo{author}{\bibfnamefont{W.~G.} \bibnamefont{Unruh}} \bibnamefont{and}
  \bibinfo{author}{\bibfnamefont{R.}~\bibnamefont{Schutzhold}},
  \bibinfo{journal}{Phys. Rev.} \textbf{\bibinfo{volume}{D71}},
  \bibinfo{pages}{024028} (\bibinfo{year}{2005}), \eprint{arXiv:gr-qc/0408009}.

\end{thebibliography}
\end{document}